\documentclass[intlimits,twoside,a4paper]{article}

\usepackage[cp1251]{inputenc}
\usepackage[eqsecnum]{cmpj3}
\usepackage{bm}
\usepackage{graphicx}
\usepackage{latexsym}
\usepackage{color}

\def\be{\begin{equation} }  
\def\ee{\end{equation} }

\def\iD{{\mbox{\sc \footnotesize d}}}
\def\iPY{{\mbox{\sc \scriptsize py}}}
\def\iBM{{\mbox{\sc \scriptsize bm}}}

\def\iTPT{{\mbox{\rm \tiny TPT}}}
\def\iSPT{{\mbox{\rm \tiny SPT}}}
\def\iISPT{{\mbox{\rm \tiny ISPT}}}
\def\iSFP{{\mbox{\rm \tiny SFP}}}
\def\iISFP{{\mbox{\rm \tiny ISFP}}}

\def\iNew{{\mbox{\rm \scriptsize New}}}

\issue{2021}{24}{3}{33504}
\doinumber{10.5488/CMP.24.33504}
\title[Relationship between TPT and SPT for fused dimers fluids]%
{Relationship between thermodynamic perturbation and scaled particle theories for fused dimers fluids %
}
\author[O. Bernard]{O. Bernard}
\address{
 Laboratoire PHENIX, CNRS, Sorbonne Universit\'e (Campus P.M. Curie), 4 Place Jussieu, \\Case 51, F-75005 Paris, France}

\date{Received July 6, 2021, in final form August 2, 2021}

\begin{document}

\maketitle

\begin{abstract}

Various approaches 
 are reviewed that use 
 scaled particle theories to describe dumbbell fluids
made of tangent or overlapped hard spheres. 
Expressions encountered in the literature are written  
 in a form similar to that presented in the thermodynamic 
perturbation theory introduced by Wertheim for chains and developed in statistical associating fluid theory (SAFT). 
Analogies and differences observed in these two types of theoretical descriptions allow one to  propose alternative 
theoretical expressions to describe dumbbell fluids with overlapping spheres.


\end{abstract}

\section{Introduction}

The description of thermodynamic properties of media containing non-spherical objects is still important 
in many fields of fundamental or applied science. It still presents difficulties, in particular in the 
description of 
phase transitions and the various types of phases generated according to the geometry of the objects.   
 Convex objects were described previously using  extensions of scaled particle theory (SPT), 
 which was initially introduced for spheres~\cite{Helfand60}.   
 These extensions, sometimes found  using virial developments, have given rise to various conformal 
 theories~\cite{Boublik75,Barboy79,Rosenfeld88,Carnahan99}. They   
have led in particular to free volume theories~\cite{Oversteegen05} 
used to describe phase transitions, as well as to some  approaches of inhomogeneous fluids~\cite{Rosenfeld89}.

 Furthermore, the chains of spheres linked together could also be described within the thermodynamic 
 perturbation theory (TPT) proposed by Wertheim~\cite{Wertheim84a,Wertheim84b}, 
 which led to the statistical associated fluid theory~(SAFT)~\cite{Chapman89} and its extensions currently used to describe polymer fluids. 
 This  approach separates the thermodynamic quantities considered (free energy or pressure) 
 into a reference contribution, related to the subunits in absence of connection,  
 and into a contribution induced by the connectivity between these subunits 
 necessary to form the chains.   
 In addition, the spheres constituting the chains are most often constrained to be tangent because the description of spheres
with overlap proved to be more difficult. 
 However, in the extensions of the TPT used to describe polymers, it is often useful to be able to describe various subunits within chains as groups of spheres with overlap.
  Better consideration of such subunits with the scaled particle theory could be useful to improve the application of TPT to polymers.  

 On the other hand,  chains of spheres cannot in principle be described with scaled particle theory 
 or its variants because these objects are non-convex.
However, the theoretical SPT models of spherocylinders have been adapted to describe linear chains.
In particular, this has provided a good description of dumbbell fluids formed of two overlapping spheres.  
This is probably due to the similarity in shape between these dumbbells and slightly elongated spherocylinders. 
  Another conformal theory, called scaled field particle (SFP) theory and having similarities with SPT,  
 has also been used to describe dimer fluids.
 Then, in this work, these two versions of conformal theories were reviewed and applied to describe thermodynamics of dimer fluids.    
  This paper is dedicated to Yu.~V.~Kalyuzhnyi, who has made significant contributions in many different research areas related to statistical mechanics, 
especially in integral equation theories.


  The reference contribution in TPT is that of dissociated hard spheres.       
 Then, in the next section,  
   the description of hard spheres and its use in TPT theory is presented.    
 A theoretical expression of the dimer pressure obtained using integral equations is also given to later 
 highlight links with the other approaches presented. 
 In the following section, various conformal or free volume theories inspired by SPT are presented.  
  Conformal theories describe the thermodynamic quantities of fluids by using geometric   
 parameters characteristic of the bodies considered, such as their volumes, surface areas and   
 mean radii. The two types of conformal theories studied do not use exactly the same set
 of parameters.   
  In addition, for each type of conformal theory, from the original result a corrective term has been 
 added to obtain an improved version. The corrective term  depends also on the set of geometric parameters. 
 Since the two types of conformal theories do not use exactly the same geometric parameters, the corrective term is different.  
 So, for each theory we have presented both the original version and the improved version to be able to better study the influence
of the corrective term. 
   
  At small separation between the spheres constituting the dimer, when the larger sphere tends to completely 
  overlap the smaller one, it is known that the two types of conformal theories tend towards the same result. 
  The difference between the results given by the two approaches increases with the elongation between the 
  components of the dimer. 
   Then, in the next section, in order to highlight the differences between the two conformal approaches,   
  these theories were first applied to dimers made of tangent spheres. For this particular  
  configuration, all of the above-mentioned geometric parameters can be expressed from the diameters of the hard spheres.  
  Then, the thermodynamic quantities coming from the various conformal theories, were rewritten as a function of the diameters of the hard
 spheres instead of the aforementioned geometric parameters. In particular, when the hard spheres have all the same diameter, the pressure 
 of the dimer fluid is described simply as a function of the volume fraction.  
 Moreover, these expressions were split into two part according to the same dividing found in the TPT, 
 namely a reference term of dissociated hard spheres and a connectivity term due to the links between spheres. 
 This made it possible to better analyze the reference term and the connectivity term in each of the theories.  
 This made it possible to explain the differences observed in the description of the thermodynamics deduced from simulations.  
 
 After this analysis of previous works, in the following section alternative relationships have been established to describe 
 first dumbbells made of tangent spheres, and then those with overlap. A comparison was given with the results of simulations.      
 In a last part we conclude this work and present the prospects that it suggests.

\section{From integral equations to perturbation theory for hard spheres and  tangent hard sphere dimers}

\subsection{Integral equation for dissociated hard spheres fluid} 

 A proper account of the excluded volume of various species is essential to describe 
 thermodynamic properties such as the pressure $P$ 
  of liquids and 
  concentrated solutions~\cite{Barker67a,Barker67b,WCA1,WCA2}.    
Since the sphere is the simplest object taking into account the repulsion at short distances, 
 all the constituents were first considered as hard spheres. 
 In addition, molecules of complex shape can be represented using collections of spheres 
 linked together. Therefore, it seems essential to first properly describe the 
 properties of these subunits constituting these assemblies. Numerical simulations make it 
 possible to precisely determine the structural and thermodynamic properties of model systems. 
 However, their implementation is cumbersome and does not lead to explicit expressions of the 
 studied quantities.   
  As an alternative, studies of the structural 
and thermodynamic properties of hard spheres  
 were undertaken using the Ornstein-Zernike (OZ) integral equations~\cite{OZ14}.  
 Among various closure relations useful in solving these integral equations, 
 the Percus Yevick (PY) approximation~\cite{PY58} has proven particularly suitable to describe hard spheres.
In addition, it provides analytical expressions of some structural and thermodynamic quantities.  
Notably, the PY approximation allows one to access the radial distribution functions $g_{ij}(r)$ between 
 all pairs of components $i$ and $j$ present in the liquid~\cite{Wertheim63,Tiele63,Lebowitz64}.  
 Various thermodynamic quantities can be calculated from the radial distribution functions $g_{ij} (r)$~\cite{Hansen}.   
 In particular, from the relation defining the compressibility as a function of these $g_{ij}(r)$, 
 an analytic expression of the pressure has been deduced by integration~\cite{Lebowitz64b,Baxter70,Salacuse82} 
 \be 
\beta  P_{\iPY}^{c} = \frac{6}{\piup} \left[ \frac{\zeta_0}{1 - \zeta_3} + \frac{3 \,  \zeta_1 \zeta_2}{\left( 1 - \zeta_3 \right)^2} 
+ \frac{3  \, \zeta_2^3}{ \left( 1 - \zeta_3 \right)^3 } \right], 
\label{PYCP}
\ee   
where $\beta$ is equal to $1 / k_{\rm B} T$ with $k_{\rm  B}$ the Boltzmann constant and $T$ the temperature.   
 The terms $\zeta_n$  are given by
\be
\zeta_n = \frac{\piup}{6} \sum_k \rho_k \sigma_k^n,
\ee
where $\rho_k$ and $\sigma_k$ are  the number density and the diameter of component $k$, respectively. 
  Subscript ``{\scriptsize PY}'', under the symbol $P$, means that the pressure was obtained in the PY approximation, 
and the superscript ``$c$'' has been added to remind that this expression comes from the compressibility relation.  
 The virial relation also provides an analytical expression of the pressure $P_{\iPY}^v$, 
 which is however less precise in comparison with the results 
  deduced from numerical simulations.  
  An alternative expression  of the pressure somewhat more accurate than that given by equation~(\ref{PYCP}), 
	has been deduced by a combinaison 
 of the pressure deduced from the virial and the compressibility relations, first in the case of a single component~\cite{Carnahan69} 
  and then for mixtures of several constituents~\cite{Boublik70,Mansoori71} : 
  \be 
\beta P_{\iBM} =  \frac{6}{\piup} \left[ \frac{\zeta_0}{1 - \zeta_3} + \frac{3 \, \zeta_1 \zeta_2}{\left( 1 - \zeta_3 \right)^2} 
+ \frac{3 \, \zeta_2^3 - \zeta_2^3 \zeta_3}{ \left( 1 - \zeta_3 \right)^3 } \right].
\label{BMP}
\ee 
 The subscript ``{\scriptsize BM}'' means Boubl\'{\i}k-Mansoori, to recall the initials of the  
 two first authors of the articles in which were established the corresponding expression 
 for mixtures of several constituents~\cite{Boublik70,Mansoori71}.   
The ``{\scriptsize BMCSL}'' subscript  (or superscript) is often used to denote this equation, to recall 
the initials of all the  authors who established it for mixture of components~\cite{Boublik70,Mansoori71}.  Here, 
 we limited the number of letters in subscript to alleviate the notation.  
 Expressions giving an even better description of the pressure obtained by simulation were proposed later.
 However, in this article we limit ourselves to this level of refinement, 
 compared to the expression~(\ref{PYCP}) determined using the PY approximation. 
 Indeed, the passage of equation~(\ref{PYCP}) to~(\ref{BMP}) is simple enough to 
 be used later in the context of expressions describing non-spherical objects.   
 Moreover, equation~(\ref{BMP}) is commonly used both in perturbation theories and 
  in applications to describe experimental systems. 
  Thereafter, from these expressions of the pressure,  other quantities, such as the free energy $A$ 
 or the chemical potentials $\mu_j$ of  various $j$ constituents, can be deduced by integration 
 or differentiation.  They will also be useful when applying perturbation 
 theories which take the fluid of dissociated hard spheres as a reference.

\subsection{Integral equations for dimers made of  tangent hard spheres} 
 
 \indent Model molecules, made of assemblies of hard spheres were considered.
  Dimers made of hard tangent spheres are the simplest of these aggregates.  
  In order to describe dimers with integral equations, 
  attractive sites were added on their surfaces.  These sites induce the formation of bonds between 
    the spheres.   
If there is only one site per sphere,  only dimers can be formed.    
 Since the sites occupying a very small fraction of the surface
of the spheres, the attractive interactions are  highly directional. 
Dimers are formed only when two contacting spheres have their sites oriented towards each other.  
In the other configurations,  particles interact only with the purely radial potential 
 (hard spheres potential).   
 The consideration of these highly directional potentials permits using  
 various integral equations   
 to describe the structure and interactions of these hard sphere assemblies~\cite{Chandler82,Rossky84,Wertheim84a,Wertheim84b,Wertheim86a,Wertheim86b,Wertheim86c,Wertheim87,Kalyuzhnyi93a}.   
 In particular, Wertheim's OZ equations (WOZ)~\cite{Wertheim84a,Wertheim84b,Wertheim86a,Wertheim86b,Wertheim86c,Wertheim87}
 were applied to the description of dimers formed by tangent spheres~\cite{Wertheim86c,Yakub92,Kalyuzhnyi93b,Kalyuzhnyi94}.  
	As an extension of the PY relation, a polymer Percus Yevick (PPY) closure relation was used.    
Moreover, the density $\rho_i$ of various $i$ species, is decomposed into the density of 
 free $i$ species $\rho_i^0$, and dimers $\rho_{\iD}$. 
 Considering, for simplicity only, two species $1$ and $2$ which can link together to form dimers, 
   we have the mass conservation relations,    
  for $i = 1$ or $2$, 
	\be
\rho_i = \rho^0_i + \rho_{\iD},
\label{MAL1}
\ee
with
\be
\rho_{\iD } = \rho_1^0 \; \rho_2^0 \; K^0 \; g^{00}_{12}(\sigma_{12}),
\label{rhoD}
\ee
where $K^0$ is an association constant (which can be expressed as an integral dependent on the attractive potential~\cite{Wertheim86c}), 
 and $g^{00}_{12}(\sigma_{12})$  is the pair distribution  between unassociated species 
 $1$ and $2$ at a contact distance $\sigma_{12} = (\sigma_1 + \sigma_2)/2$.   
 The solution of the WOZ equations allows one to accurately describe the interactions between 
 these constituents. 
 As in the case of the dissociated hard spheres, analytic expressions of the pressure, deduced from 
 the virial and compressibility relations, were obtained. 
 Here again, the expression deduced from the compressibility is the most efficient~\cite{Wertheim86c,Kalyuzhnyi94}.  
 \begin{subequations}
	\begin{align}
		\beta  P_{\mbox{\rm \tiny PPY}}^{c} & = \beta  P_{\mbox{\rm \tiny PY}}^{c} -  \rho^0_1 \rho^0_2 \; K^0 \Bigl[ g_{12}^{00}(\sigma_{12}) \Bigr]^2 
		\label{PPYa}  \\ 
		& = \beta  P_{\mbox{\rm \tiny PY}}^{c} - \rho_{\mbox{\sc \footnotesize d}} \; g_{12}^{00}(\sigma_{12})  .
		\label{PPYb} 
	\end{align}
\end{subequations}
 The definition of $\rho_{\iD}$, given by the equation (\ref{rhoD}), has been used   
to go from equation~(\ref{PPYa}) to~(\ref{PPYb}).  Furthermore,    
 in the PPY approximation, it was found  that    $g_{12}^{00}(\sigma_{12})$  is equal 
 to  $g_{12}^{\mbox{\rm \tiny PY}}(\sigma_{12})$~\cite{Wertheim86c,Kalyuzhnyi93b},   
which is the corresponding pair distribution function at contact, 
 obtained in the PY approximation for hard spheres without attractive sites. 
Moreover, $g_{12}^{\mbox{\rm \tiny PY}}(\sigma_{12})$  is given by~\cite{Lebowitz64}   
\be 
 g_{12}^{\mbox{\rm \tiny PY}}(\sigma_{12}) = 
\frac{1}{1- \zeta_3} +  \frac{ 3 \, \zeta_2}{\left(1 - \zeta_3\right)^2} \, \frac{\sigma_1 \sigma_2}{\sigma_1 + \sigma_2}.
\label{gPY} 
\ee
Equation~(\ref{PPYb}) was also established within various integral equations devoted to the description of properties of chain fluids~\cite{Chiew90,Kalyuzhnyi96,Stell99a,Stell99b}.    
As a generalisation in equation~(\ref{PPYb}), for chains of density $\rho_c$ containing $m$ subunits,  
 $\rho_{\mbox{\sc \footnotesize  d}}$ is replaced by 
$(m-1) \, \rho_c$.    
In the next section, when comparing the various approaches, 
 we come back to the explicit expression~(\ref{PPYb}) derived from the compressibility relation   
within the PPY approximation. 

\bigskip
\subsection{Thermodynamic perturbation theory for dimers made of  tangent hard spheres} 

As an alternative, a thermodynamic perturbation theory (TPT)  was also developed to take 
 into account the effect of the bonds formed between the particles on the thermodynamic 
 properties of the fluid~\cite{Wertheim84a,Wertheim84b,Wertheim86a,Wertheim86b,Wertheim86c,Wertheim87}.  
 The  generalization  and application of this theoretical path in  the SAFT led to   
  one of the most accurate equations of state for tangent hard sphere chains~\cite{Chapman89,Jackson88,Chapman88,Archer91}.  These  approaches describe the thermodynamics 
  in terms of the reference fluid distribution functions regardless of the amount of bonding. The residual Helmholtz 
  free energy $A$ can be separated into two terms as follows: 
\be
  \frac{\beta A_{\iTPT}}{V}   =   \frac{\beta A^{\rm ref}}{V}  +\frac{\beta A^{\rm bond}}{V}, 
\ee
where $A^{\rm ref}$ is the reference  Helmholtz free energy contribution from the dissociated free monomers, 
 and $A^{\rm bond}$ is a contribution related to the formation of bonds.
In the 1st order perturbation theory, the term $A^{\rm bond}$ can also be evaluated
from the properties of the reference system without association. 
\be
\frac{\beta A_{\iTPT}}{V}  
    =   \frac{\beta A^{\rm ref}}{V}  - (m -1) \, \rho_c \ln{ g^{\rm ref}(\sigma_{12})}.
  \label{TPTA}
\ee
In the second term of equation~(\ref{TPTA}), related to the connectivity within the chains,  
  $m$ is the number of monomers, $\rho_c$ the number density of the chain and 
 $g^{\rm ref}(\sigma_{12})$   the pair distribution function between two monomers at a distance of contact  
  $\sigma_{12}$ in the reference fluid without association. 
	For simplicity, it is assumed here  that all the monomers 
  are separated by the same  contact distance in the chains.  
  If the reference fluid consists of hard spheres, the reference free energy can be described 
	using either the PY expression, 
  \be
\frac{\beta A^{c}_{\iPY}}{V} = \frac{6}{\piup} \left[ - \zeta_0 \ln{(1 - \zeta_3)} + \frac{3 \,  \zeta_1 \zeta_2}{1 - \zeta_3} 
+ \frac{3  \, \zeta_2^3}{2 \left( 1 - \zeta_3 \right)^2 } \right],
\label{APY}
\ee
  corresponding to the expression of the pressure given by equation~(\ref{PYCP}) 
    or  the improved expression associated with the pressure given by~(\ref{BMP}), 
 \be
\frac{\beta A_{\iBM}}{V} = \frac{6}{\piup} \left[ \left( \frac{\zeta_2^3}{\zeta_3^2} - \zeta_0 \right) \ln{\left( 1 - \zeta_3 \right)} 
 + \frac{3  \, \zeta_1 \zeta_2}{1 - \zeta_3} + \frac{\zeta_2^3}{\zeta_3 \left( 1 - \zeta_3 \right)^2}  \right].
 \label{ACS}
\ee   
 This latter expression is the one generally used because it leads to the results closest to those 
 given by the simulations, in comparison with those given by equation~(\ref{APY}). 
 However, we have also presented expression~(\ref{APY}), deduced from the PY theory, 
 because it can also be recovered using other theories 
 presented below in this article. This property will be used later to develop new improved expressions.    
In the second term of equation~(\ref{TPTA}), the pair distribution function at contact can be described 
either using an expression deduced from the OZ equations, or using an expression corresponding to the improved expression of pressure~(\ref{BMP}).  
The pressure $P_{\iTPT}= P^{\rm ref} + P^{\rm bond}$ can be deduced from equation~(\ref{TPTA}) by differentiation 
\be
  \beta P_{\iTPT} =   \beta P^{\rm ref}  - (m -1) \, \rho_c \left[ 1 + \sum_k \rho_k \frac{\partial \ln{ g^{\rm ref}(\sigma_{12})}}{\partial \rho_k} \right].
  \label{TPTP}
  \ee
  In the same way, when the reference fluid consists of hard spheres,  the reference pressure $P^{\rm ref}$ can be described either 
  using the PY expression (\ref{PYCP}), or the improved expression (\ref{BMP}).  It is recalled that 
   in the case where only dimers are formed, $m=2$, the concentration of chains $\rho_c$ 
	 denotes the concentration of dimers    $\rho_{\iD}$.  
	 Then, in equations~(\ref{TPTA}) and~(\ref{TPTP}), $(m-1) \rho_c$ can be simply replaced by $\rho_{\iD}$. 
 
 
 Attempts have been made to extend TPT to fluids of fused sphere chains~\cite{Boublik89,Walsh90,Boublik90,Amos92}.  
 In equations~(\ref{TPTA}) and~(\ref{TPTP}), the distribution function $g^{\rm ref}(\sigma_{12})$ is replaced by a cavity correlation function $y (l)$ where 
 $ l \leqslant \sigma_{12}$, is  the distance between the spheres linked together.    
 Moreover, an effective number of tangent monomers was used by requiring that the second virial coefficient of the chain of tangent spheres should be equal to 
   that of  the chain of fused spheres.  In the case of fused spheres, this second virial coefficient can be expressed in terms of  parameters of 
   non-sphericity which account for the molecular shape of the chain. This approach is based on the contribution of various conformal theories describing 
   the properties of objects of various shapes.  
   
   \section{Conformal or free volume theories}
   
  One of the goals of conformal or free volume  theories is to attempt to describe individual chemical potentials and an equation of state of convex body  in terms of a relevant 
  set of parameters such as the volume, surface area and the
mean radius of curvature which characterize the shape of all the components.   
  This approach was first performed in the scaled particle theory (SPT). 

\subsection{Scaled particle theory} 
   
The SPT was first developed to describe the hard sphere fluids~\cite{Helfand60,Lebowitz65}.  
Analytical expressions of the radial distribution functions at  contact and of thermodynamic 
 quantities were obtained.  
It is noted, that the resulting expression of the pressure is identical to that given in equation~(\ref{PYCP}), 
 derived from the compressibility relation and the PY approximation.  
The SPT was extended to the description of  convex objects other than hard spheres, 
such as ellipsoids and spherocylinders.  
 In particular, the studies of prolate spherocylinders led to an analytical description of 
 the pressure~\cite{Gibbons69,Gibbons70,Boublik74} 
 \be
\beta P_{\iSPT} = \frac{\rho}{1-v} + \frac{r s}{\left(1 - v \right)^2} + \frac{q s^2}{3 \left( 
 1 - v \right)^3},
\label{SPTP}
\ee  
where:
\[
\rho = \! \sum_i  \rho_i, \;  \;  v =  \! \sum_i \rho_i V_i,  \; \;  s = \! \sum_i \rho_i S_i,  \;  \; r =  \! \sum_i \rho_i R_i  \;  \; 
 \mbox{and} \;  \; q = \! \sum_i \rho_i R_i^2.
 \label{shape} 
\] 

\noindent This result  for $P$,  is no longer  only  
a function of the diameters of the spheres,  but is also expressed as a function of the volume $V_i$, 
of the surface area $S_i$ and of the mean curvature $R_i$ of each $i$  object. 
This formula is a generalization of that obtained within SPT for mixtures of hard spheres. 
In this case, the geometric parameters are related to the diameters $\sigma_i$:   
 $R_i = \sigma_i / 2$, $S_i = \piup \sigma_i^2$ and $V_i =  (\piup / 6) \sigma_i^3$. 
Using these relations in equation~(\ref{SPTP}), leds to the relation~(\ref{PYCP}), previously 
 deduced from the compressibility in the PY approximation.        


An improved scale particle theory (ISPT) was developed as a generalization of 
 equation (\ref{BMP}),   
 and a more accurate expression of the pressure was obtained~\cite{Boublik75}   
\be
\beta P_{\iISPT} =   \frac{\rho}{1-v} + \frac{r s}{\left(1 - v \right)^2} + 
 \frac{q s^2 \left( 3 - v \right)}{9 \left( 1 - v \right)^3} .
 \label{ISPTP} 
\ee
Here also, when all the constituents are spherical, this formula becomes identical to that given by equation~(\ref{BMP}).    
 In principle, SPT is only applicable to convex objects. 
 Dumbbells  resemble prolate spherocylinders but are not convex and the result of the SPT should not 
 be used for these model molecules.   Nevertheless, the previous equation~(\ref{ISPTP}) was applied  to fused dimers~\cite{Boublik77} and 
 the results agree well with  the simulation data~\cite{Tildesley80}.    
 Otherwise, equation~(\ref{SPTP}) leads to an exact second virial coefficient and to an approximate third coefficient.  
 On the other hand, equation~(\ref{ISPTP})  provides a third virial coefficient closer to the exact value.  
 It is  noted that the different values found for the third coefficient seems to be related to the difference 
 observed between the third terms to the right of the  equality in each  of these two equations.   
Moreover, there are other theoretical developments attempting to jointly describe the thermodynamics 
 and the structure of fluids of convex objects. 

\subsection{Scaled field particle theory} 

Scaled field particle (SFP) theory provides an explicit expression of direct correlation functions in addition 
to the quantities also determined by SPT~\cite{Rosenfeld88}.   
The expressions of various thermodynamic quantities are very close to those deduced within the SPT. 
In particular, the pressure is written with an expression very close to equation~(\ref{SPTP}), 
 derived under the SPT: 
\be
\beta P_{\iSFP} = \frac{\rho}{1-v} + \frac{r s}{\left(1 - v \right)^2} +  \frac{1}{12 \piup} \frac{ s^3}{ \left( 
 1 - v \right)^3}.
\label{SFTP}
\ee 
This formula   differs only in the last term, where $q s^2$ is replaced by $s^3 / (4 \pi )$. 


In the same way, an improved scale field relation (ISFP) was proposed as a generalization of 
that leading to equation~(\ref{BMP}).  
This relation can be easily deduced from equation~(\ref{ISPTP}), by replacing, in the third term to the right 
of the equality, $q s^2$ by $s^3 / (4 \piup)$.  
\be
\beta P_{\iISFP} =   \frac{\rho}{1-v} + \frac{r s}{\left(1 - v \right)^2} + 
 \frac{ s^3 \left( 3 - v \right)}{36 \piup \left( 1 - v \right)^3}. 
 \label{ISFPP} 
\ee

From the SPT and the SFP, we have  introduced four possible expressions to account for the pressure variations 
observed in a fluid of  convex objects. 
In the following sections, a comparison between these various relationships is presented, when used to describe 
a fluid  of tangent or fused hard dimers.  
 New  expressions are also established based on the mathematical form of the TPT equations~(\ref{TPTA}) and~(\ref{TPTP}).

\section{Application of conformal theories to dumbbells}

 A comparison of the results obtained using various conformal theories  introduced 
in the previous section may be done, when the fluid is made of dumbbells. 
 Dumbbells made of two spheres, denoted by indices $1$ and~$2$,  of diameters 
 $\sigma_1$ and $\sigma_2$, separated by   
 a distance:  $l \leqslant \sigma_{12}  \equiv  (\sigma_1 + \sigma_2)/2$,  are considered.   
 The characteristic quantities $V_{12}, S_{12}$ and $R_{12}$ of these objects are~\cite{Rosenfeld88,Boublik90,Amos92}  
\be
V_{12} = \frac{\piup}{12} \left[ \sigma_1^3 + \sigma_2^3 + \frac{3}{2} \left( \sigma_1^2 + \sigma_2^2 
\right) \; l + \frac{3}{16} \frac{\left( \sigma_1^2- \sigma_2^2 \right)^2}{l} - \; l^3 
\right], 
\label{Vdb}
\ee
\be
S_{12} = \frac{\piup}{2} \left[ \sigma_1^2 + \sigma_2^2 + \left( \sigma_1 + \sigma_2 \right) \; 
\left( l + \frac{\left(\sigma_1 - \sigma_2 \right)^2}{4 \; l } \right) \right],
\label{Sdb}
\ee
\be
R_{12} = \frac{1}{4} \left[ \sigma_1 + \sigma_2 + \; l + \frac{\left( \sigma_1 - \sigma_2 
\right)^2}{4 \; l} \right].
\label{Rdb}
\ee

Assuming that $\sigma_1$ is greater than $\sigma_2$, one can first determine the values taken by these 
quantities when $l \leqslant  (\sigma_1 - \sigma_2) / 2$.   
In this case, sphere $2$ is completely included in sphere $1$ and various geometric quantities defined above 
take the values corresponding to sphere $1$ alone.  
 Once these quantities are known, when only dumbbells are present (with $\rho_1 = 
\rho_2$),  the pressure can be determined.  
 It is found that the expressions of $P_{\iSPT}$  and $P_{\iSFP}$ are identical to 
 $P_{\iPY}^c$,  given by equation~(\ref{PYCP}), with  $\zeta_n = (\piup/6) \rho_1 \sigma_1^n$ and 
  $n= 0,1,2$ and $3$. 
In the same way,  the expressions of $P_{\iISPT}$  and $P_{\iISFP}$ are identical to 
 $P_{\iBM}$,  given by equation~(\ref{BMP}).   
Then, when  $l \leqslant  (\sigma_1 - \sigma_2) / 2$, 
 SPT and SFP  expressions of the pressure merge to the  PY compressibility 
relation for hard spheres, and ISPT and ISFP expressions merge to the improved BMCSL for hard spheres.


The limit where $l = \sigma_{12}$ is also very important. 
Again, in this case the various geometric quantities can be explained in terms of the diameters of the spheres only. 
 The thermodynamics deduced from conformal theories leads to relations similar to those obtained with TPT. 
 These relationships are presented in the following subsections. 

\subsection{Pressure for dimers made of tangent hard spheres}

 When the two spheres are in contact: $l = \sigma_{12}$,  then   
 $V_{12} = (\piup / 6) (\sigma_1^3 + \sigma_2^3)$ and  $S_{12} = \piup (\sigma_1^2 + \sigma_2^2)$, 
  are  the sum of the volumes and surface areas of the two spheres, respectively.  
 However, the quantity $R_{12}$ is different from the sum  
 of the two separate radii, namely  $R_{12} = (\sigma_1 + \sigma_2) / 2 - \; \sigma_1 \sigma_2 
  / [2 (\sigma_1 + \sigma_2)]$. 


Now, in the case of a dimer formation 
 equilibrium, let us denote by   
 $\rho^0_1$ and $\rho^0_2$, the densities of fully dissociated particles $1$ and $2 $, which can form dumbbells D, 
of  number density $\rho_{\iD}$, with the equilibrium
\be
(1) + (2) \rightleftharpoons (\mbox{D}).   
\ee   
With  this chemical equilibrium,  it is observed that the concentrations 
that naturally occur are $\rho^0_1$, $\rho^0_2$ and  $\rho_{\iD}$ and not 
directly $\rho_1$ and $\rho_2$, the total concentrations (free or bound) of particles $1$ and $2$.  
 Then, from the SPT equations, the  
 pressure $P_{\rm SPT}$  can be evaluated by considering the fluid 
 as a mixture of two  kinds of hard spheres components, of  densities $\rho^0_1$ and $\rho^0_2$, 
  and dimers, of density $\rho_{\iD}$.  
  The geometric parameters, $R_i$, $S_i$, $V_i$, of these two kinds of an object can be determined thanks to the equations~(\ref{Vdb})--(\ref{Rdb}).  The introduction of equilibrium makes it possible to link the chemical potentials of free and linked species.   
 In addition, the quantities of the three species involved in equilibrium are related 
 by the mass conservation relations~(\ref{MAL1}).  
 It has been noted previously~\cite{Olaussen91,Fisher98} that the thermodynamic relationships 
 for systems with association can be described equivalently, 
either with the densities of the various species in equilibrium or with the total densities 
of species $1$ and $2$, considering $\rho_{\iD}$ 
the density of pairs  as a function of total densities. 
 In the case of dumbbells made of tangent spheres,
 the integral equations within the PPY approximation and the perturbation theory lead to expressions 
 which are functions of the total densities and of the density of dimers.   
 Then, from the conformal expressions~(\ref{SPTP})--(\ref{SFTP}), 
 equations are established which are also expressed as  a 
function of the total and dimer densities.  

Thus, in the  SPT equation~(\ref{SPTP}) with three constituents,  one can first replace 
 the densities $\rho^0_1$ and $\rho^0_2$ by $\rho_1 -  \rho_{\iD}$ 
  and $\rho_2 - \rho_{\iD}$, respectively, 
	in the quantities $\rho$, $r, s, q$ and $v$.  One then obtains, when $l = \sigma_{12}$ 
 \be 
 \beta P_{\iSPT} = \beta P_{\iPY}^c - \rho_{\iD} \; \left[   
 g_{12}^{\iPY}(\sigma_{12})  -  
  \frac{3 \zeta_2^2}{\left( 1 - \zeta_3 \right)^3 } \;  \frac{\sigma_1^2 \sigma_2^2}{\left( \sigma_1 + \sigma_2 \right)^2 } \right], 
  \label{SPTPS}
  \ee 
  where $g_{12}^{\iPY}(\sigma_{12})$ is given by equation~(\ref{gPY}).   
   In the equation~(\ref{SPTPS}), the terms $\zeta_n$ ($n=0,\ldots,3$), appearing in  $\beta P^c_{\iPY}$  
   and in the second part of the equation,   
  are calculated by considering the total concentrations $\rho_1$ and $\rho_2$ 
of the two constituents : 
\be
\zeta_n = \frac{ \piup}{ 6} \left( \rho_1 \sigma_1^n + \rho_2 \sigma_2^n \right).  
\label{zeta2} 
\ee

\noindent  In the same way, for the improved SPT,  equation~(\ref{ISPTP})  led to 
  \be 
 \beta P_{\iISPT} = \beta P_{\iBM} - \rho_{\iD} \; \left[  
 g_{12}^{\iPY}(\sigma_{12})  -  
  \frac{ \zeta_2^2 \left( 3 - \zeta_3 \right)}{\left( 1 - \zeta_3 \right)^3 } \;  \frac{\sigma_1^2 \sigma_2^2}{\left( \sigma_1 + \sigma_2 \right)^2 } \right]. 
  \label{ISPTPS}
  \ee 
  
  \noindent The same procedure can be applied to expressions deduced from the SFP theories.  
	 When $l = \sigma_{12}$, equation~(\ref{SFTP})   led to 
  \be 
 \beta P_{\iSFP} = \beta P_{\iPY}^c - \rho_{\iD} \;  g_{12}^{\iPY}(\sigma_{12}).
  \label{SFTPS}
  \ee 
  Similarly, from the equation  for the improved SFP theory, one gets  
  \be 
 \beta P_{\iISFP} = \beta P_{\iBM} - \rho_{\iD} \;  g_{12}^{\iPY}(\sigma_{12}).
  \label{ISFTPS}
  \ee 
  Note that the four equations~(\ref{SPTPS}), (\ref{ISPTPS}), (\ref{SFTPS}), (\ref{ISFTPS}) have a mathematical form similar to that of the TPT equation 
	given by~(\ref{TPTP}).
   \begin{figure}[!t]    
\centerline{\includegraphics[width=0.70\textwidth]{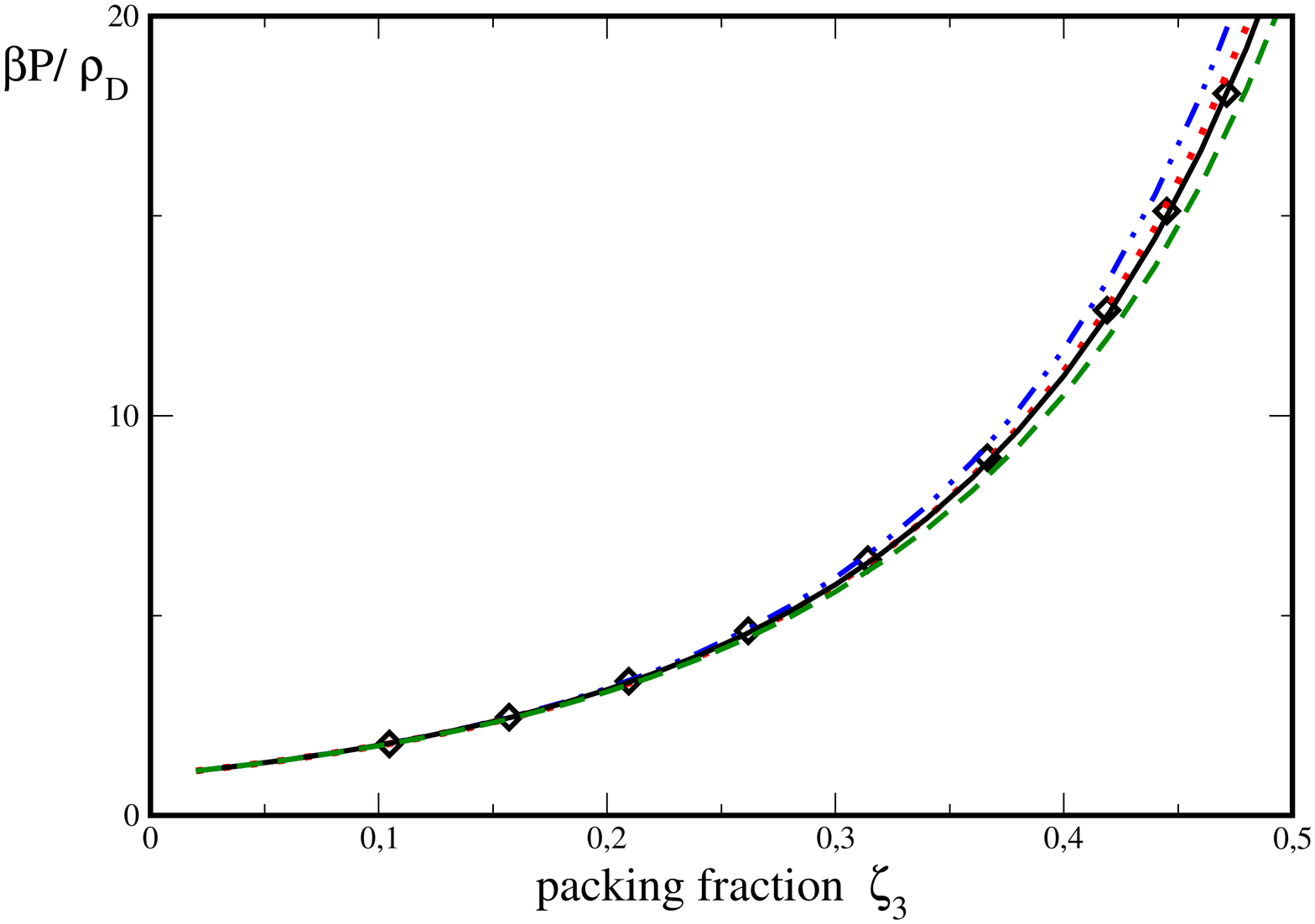}}
\caption{(Colour online) Equation of state for tangent hard sphere dumbbells. 
Lines from top to bottom represent, respectively, {
$- \cdot \cdot \, -$}
 SPT equation (\ref{SPTPS}), 
{
$\cdot  \cdot  \cdot  \, \cdot$} SFP equation~(\ref{SFTPS}),  
 --- ISPT equation (\ref{ISPTPS}), {
  -~-~-}~ISFP  equation~(\ref{ISFTPS}).  
Diamonds are simulation results (from~\cite{Tildesley80}).} \label{fig-ex1}  
\end{figure} 
 Indeed, they are all composed of a first term, related to the  dissociated hard spheres,
 and a second term due to the connectivity between the spheres.
 By analogy with the equation~(\ref{TPTP}), the first pressure term can be regarded as 
  $\beta P^{\rm ref}$, the reference pressure, 
  and the second term as $\beta P^{\rm bond}$, the contribution due to the formation of bonds.   
 Apart from the proportionality to $\rho_{\rm D}$, the connectivity term only depends on 
 the total densities of the hard spheres.
 In addition, the reference pressure is given by the PY expression  in the original  SPT and SFP theories.  
 As expected, in the equations deduced
from the improved theories, the reference pressure is given by the BM expression.  
We notice that in SPT theories the term of connectivity is different between the original 
 version and the improved one. 
  On the contrary,   
 the connectivity term is the same in the original SFP theory and in the improved one.  
 Moreover, the expression~(\ref{SFTPS}), deduced from the original SFP theory, 
 is identical to  that given by~(\ref{PPYb}), 
 which was established using integral equations.  
  Equation~(\ref{ISFTPS}) was also proposed as an alternative to equation~(\ref{SFTPS}) 
	within the framework of chain theory~\cite{Chiew90}.

 The results of the calculations carried out with these various formulae, 
for tangent homonuclear dimers (i.e., with a distance between its constituents:   $l =\sigma_{12}$), 
are shown  in figure~\ref{fig-ex1}.       
As it has already been said before, the curve deduced from the improved theory ISPT with equation~(\ref{ISPTP}) 
is closest to the simulation results. Then, by comparing the initial expressions~(\ref{SPTP}) and~(\ref{SFTP}) obtained with SPT and SFP respectively, it is seen that they give values that 
 are too high but that the latter is closer to the simulated values.  Finally, the improved version ISFP 
 gives values that are too low.  
 
\subsection{Connection with TPT deduced from free energy}

In the previous subsection, it was shown how, from the conformal equations~(\ref{SPTP})--(\ref{SFTP}),
 applied to dumbbells formed by tangent spheres, 
 the pressure can be split into a term of hard spheres repulsion  
  and a term of connectivity between these spheres.
This separation is similar to that encountered in the expression~(\ref{TPTP}) of pressure in the TPT theory.
Now, to further the analogy between the two kinds of description, we study the equations describing the free energy in conformal theories.   
The excess free energy can be deduced from the pressure by integrating the densities in each of the approximations presented above.
Thereby,  starting from equation~(\ref{SPTP}) for $P_{\iSPT}$, we find within the SPT 
\be
\beta \frac{A_{\iSPT}}{V} = - \rho \ln{\left( 1 - v \right)} + \frac{r s}{1 - v} + 
 \frac{q s^2}{6 \left( 1 - v \right)^2}.  
\label{SPTA} 
\ee
In the same way, the free energy associated with the $P_{\rm ISPT}$ pressure is given by: 
\be
\beta \frac{A_{\iISPT}}{V} = \left( \frac{q s^2}{9 v^2} - \rho \right) \ln{ \left( 1 - v \right)} 
+ \frac{r s}{1 - v} + \frac{q s^2}{9 v \left( 1 - v \right)^2}.
\label{ISPTA} 
\ee
  As previously, the free energies $A_{\iSFP}$, related to the pressure $P_{\iSFP}$, and $A_{\iISFP}$, related to the pressure $P_{\iISFP}$, 
  can be deduced from the two preceding  equations by replacing everywhere 
  $q s^2$ by $s^3 /(4 \piup)$.   
  

Now, the explicit relations of free energy, devoted to dimers formed of tangent spheres are presented. 
 The mixture of dimers of density $\rho_{\iD}$ and of the two dissociated monomers of density 
  $\rho^0_1$ and $\rho^0_2$ is consider again. 
As previously, using mass conservation, $\rho^0_1$ and $\rho^0_2$ are expressed as a function 
of the total  concentrations $\rho_1$ and~$\rho_2$ and  of the dimer concentration, namely, $\rho^0_1 
 = \rho_1 - \rho_{\iD}$ and $\rho^0_2 = \rho_2 - \rho_{\iD}$.   


Equations~(\ref{SPTA}) and~(\ref{ISPTA}) of the SPT theory are considered first.   
From the original relation (\ref{SPTA}), when  $l = \sigma _{12}$, the application of the mass conservation 
leads to 
\be
\beta \frac{A_{\iSPT}( l = \sigma_{12})}{V} = \beta \frac{A^c_{\iPY}}{V} -\rho_{\iD} \ln{{\cal G}_{\iSPT}(\sigma_{12})},    
\label{ASPTS}
\ee
where $A^c_{\iPY}$ is given by the equation~(\ref{APY}) and 
\be
 \ln{{\cal G}_{\iSPT}(\sigma_{12})}  = - \ln{(1 - \zeta_3)} +   \frac{ 3 \; \zeta_2}{1 - \zeta_3} \;  \frac{\sigma_1 \sigma_2}{\sigma_1 + \sigma_2} 
  - \frac{ 3 \; \zeta_2^2}{2 \left( 1 - \zeta_3 \right)^2} \;  \frac{\sigma_1^2 \sigma_2^2}{\left( \sigma_1 + \sigma_2 \right)^2}. 
 \label{LGS}
 \ee  
As previously, in the expression of $A^c_{\rm PY}$  and $ \ln{{\cal G}_{\iSPT}(\sigma_{12})}$, appearing in equations~(\ref{ASPTS}) and (\ref{LGS}), 
the terms $\zeta_n$ (with $n=0,\ldots,3$) are calculated with equation~(\ref{zeta2}) by considering only  the total concentrations $\rho_1$ and $\rho_2$ 
of the two constituents.    
 A similarity can be seen between the form of equation~(\ref{ASPTS}) and that  of  equation~(\ref{TPTA}), 
 describing $A_{\iTPT}$, the TPT excess free energy. 
The first term, to the right of the equality in the equation~(\ref{ASPTS}), corresponds to the reference 
 free energy $A^{\rm ref}$  in the equation~(\ref{TPTA}),  
when the PY approximation is chosen as the reference. 
Then, the second term in equation~(\ref{ASPTS}), corresponds to the contribution due to the connectivity 
 between the hard spheres. 
We recall that in the TPT equation, the second term, proportional to  the density of dimers $\rho_{\iD}$, 
 otherwise,  depends only 
on the densities of dissociated spheres.  We observe that the second term of the equation presents the same characteristics.   
If the analogy with the TPT equation is relevant, then the second term depends on $\ln{g^{\rm ref}(\sigma_{12})}$.  
We considered that the second term of the equation~(\ref{ASPTS}) had this meaning. That is why we gave it the name 
in relation $\ln{{\cal G}(\sigma_{12})}$. Furthermore, within the framework of the SPT approximation applied to hard spheres, 
an explicit expression of the pair distribution function at contact, that we named $ g^{\iSPT}_{12}(\sigma_{12})$, 
was deduced previously~\cite{Lebowitz64b}.  
\be 
g_{12}^{\iSPT}(\sigma_{12}) = 
\frac{1}{1- \zeta_3} 
+  \frac{ 3 \, \zeta_2}{\left(1 - \zeta_3\right)^2} \, \frac{\sigma_1 \sigma_2}{\sigma_1 + \sigma_2}
+  \frac{ 3 \, \zeta_2^2}{\left(1 - \zeta_3\right)^3} \, \frac{\sigma_1^2 \sigma_2^2}
{\left( \sigma_1 + \sigma_2 \right)^2 }.
\label{gSPT} 
\ee
Then, the  logarithm of $g_{12}^{\iSPT}(\sigma_{12})$  can be written as
\be 
\ln{\left[ g_{12}^{\iSPT}(\sigma_{12}) \right] }= - \ln{ (1- \zeta_3 ) } 
+ \ln{ \left[ 1 +  \frac{ 3 \, \zeta_2}{\left(1 - \zeta_3\right)} \, \frac{\sigma_1 \sigma_2}{\sigma_1 + \sigma_2}
+  \frac{ 3 \, \zeta_2^2}{\left(1 - \zeta_3\right)^2} \, \frac{\sigma_1^2 \sigma_2^2}
{\left( \sigma_1 + \sigma_2 \right)^2 }  \right] }.
\label{lngSPT} 
\ee
It was found that equation~(\ref{LGS}) for $\ln{{\cal G}_{\iSPT}(\sigma_{12})}$  
  can be recovered by taking 
  a limited development to the second order in $\zeta_2$  of this expression.   
  

	\noindent Next, by applying the same process to the free energy $A_{\iISPT}$ of the improved SPT, 
	when $l = \sigma_{12}$, it was found 
  \be
\beta \frac{A_{\iISPT}( l = \sigma_{12})}{V} = \beta \frac{A_{\iBM}}{V} -\rho_{\iD} \ln{{\cal G}_{\iISPT}(\sigma_{12})}  
\label{AISPTS}
\ee
with
\be
 \ln{{\cal G}_{\iISPT}(\sigma_{12})}  = - \ln{(1 - \zeta_3)} +   \frac{ 3 \; \zeta_2}{1 - \zeta_3} \;  \frac{\sigma_1 \sigma_2}{\sigma_1 + \sigma_2} 
 - \left( \ln{(1 - \zeta_3)} + \frac{\zeta_3}{\left( 1 - \zeta_3 \right)^2} \right) \frac{\zeta_2^2}{\zeta_3^2}  \;  
  \frac{\sigma_1^2 \sigma_2^2}{\left( \sigma_1 + \sigma_2 \right)^2}.   
 \label{LGISP}
 \ee
 The first term, to the right of the equality in the equation~(\ref{AISPTS}), corresponds to the reference free energy~$A^{\rm ref}$  in the equation~(\ref{TPTA}),  
when the BM approximation is chosen as the reference. 


 As an alternative, 
 the same process can be applied to the free energy expressions of the SFP theory.    
From the original expression~(\ref{SPTA}),  by replacing  $q s^2$ by $s^3 /(4 \pi)$, 
when  $l = \sigma _{12}$, one gets   
\be  
\beta \frac{A_{\iSFP}( l = \sigma_{12}) }{V} = \beta \frac{A^c_{\iPY}}{V} -\rho_{\rm D} \ln{{\cal G}_{\iSFP}(\sigma_{12})},  
\label{ASFTS}
\ee
where $\ln{{\cal G}_{\iSFP}(\sigma_{12})}$ is given by  
\be
 \ln{{\cal G}_{\iSFP}(\sigma_{12})}  = - \ln{(1 - \zeta_3)} +   \frac{ 3 \; \zeta_2}{1 - \zeta_3} \;  \frac{\sigma_1 \sigma_2}{\sigma_1 + \sigma_2}.  
 \label{LGISF}
 \ee
  Now, the same type of comparison can also be made between the original SFP and 
   the  improved SFP equation.   
   By following again the same process,   
	starting from equation~(\ref{AISPTS}) [but  replacing $q s^2$ by $s^3/(4 \piup)$],  
	when $l = \sigma_{12}$, it was found   
 \be
\beta \frac{A_{\iISFP}( l = \sigma_{12})}{V} = \beta \frac{A_{\iBM}}{V} -\rho_{\rm D} \ln{{\cal G}_{\iSFP}(\sigma_{12})}.  
\label{AISFTS}
\ee
As in the case of the SPT theory, during the transition from the original SFP expression to the improved one, 
 it is seen that, in the first term, the free energy $A_{\rm PY}^c$ coming from the PY approximation is replaced by the improved~$A_{\rm BM}$ equation. 
On the other hand, contrary to what had been found in the SPT theory, 
 the second term remains unchanged during the transition from the original SFP relation to the improved one.   
 In view of the numerical results presented in figure~\ref{fig-ex1}, 
 the  expressions~(\ref{AISPTS}) of the free energy lead by differentiation   
 to a value of the pressure $P_{\rm ISPT}$ that is much better than the one deduced from equation~(\ref{AISFTS}) in the ISFP theory.   
 Since the first  term  $A_{\rm BM}$ due to the dissociated hard spheres is the same, in equations~(\ref{AISPTS}) and~(\ref{AISFTS}),  
  there is a reason to believe that the difference observed in the resulting pressures  
	comes from the connectivity terms $\ln{{\cal G}}(\sigma_{12})$.  	
	 A comparison between these various expressions of $\ln{{\cal G}}(\sigma_{12})$   
	as a function of the volume fraction  is presented in table~\ref{tbl-ex1}.  
 It is observed that equations~(\ref{LGS})  and~(\ref{LGISP}) give results of the same order 
of magnitude for the term $\ln{{\cal G}}(\sigma_{12})$, 
 unlike the SFP equation~(\ref{LGISF}).   In addition,  
 equations~(\ref{LGS})  and~(\ref{LGISP}), for $ \ln{{\cal G}_{\iSPT}(\sigma_{12})}$ and $ \ln{{\cal G}_{\iISPT}(\sigma_{12})} $,  respectively,   
  give the values very close to those for $\ln{g_{12}^{\rm SPT}}(\sigma_{12})$, 
  deduced from  equation~(\ref{gSPT}). Equation~(\ref{LGISP}) gives the closest values.   
  Then,  by comparing the improved ISPT expression  
 to the original SPT equation, 
 taking into account the fact that the term $\ln{{\cal G}}(\sigma)$ related to 
   connectivity is of the same order of magnitude in the both cases, 
	 we deduce that the main correction incorporated in equation~(\ref{AISPTS}), 
	 with respect to equation~(\ref{ASPTS}), is made to the reference term of dissociated hard spheres.  
\begin{table}[htb]
\caption{\label{tbl-ex1} Comparison between various expressions 
 of $\ln{{\cal G}}(\sigma)$, given by equations  (\ref{LGS}), (\ref{LGISP}) 
 and~(\ref{LGISF}), with $\ln{ g_{12}^{\iSPT}}(\sigma_{12})$ deduced from (\ref{gSPT}),  as a function of the volume fraction $\zeta_3$, 
 when $\sigma_1 = \sigma_2$.}  
\vskip 0.5cm
\begin{center}
\begin{tabular}{|c|c|c|c|c|}
\hline
  & & & & \\
$\zeta_3$ & $\ln{{\cal G}_{\iSPT}}(\sigma)$   & $\ln{{\cal G}_{\iISPT}}(\sigma)$ 
 & $ \ln{ g^{\iSPT}_{12} }(\sigma)$ & $\ln{{\cal G}_{\iSFP}}(\sigma)$ \\
\hline
\hline
0.05 & 0.1292 & 0.1292 & 0.1292 & 0.1302 \\
\hline
0.1 & 0.2674  & 0.2675  & 0.2674 & 0.2720 \\
\hline
0.15 & 0.4155 & 0.4160 & 0.4157 & 0.4272 \\
\hline 
0.2 & 0.5747  & 0.5758 & 0.5751 & 0.5981 \\
\hline
0.25 & 0.7460 & 0.7485 & 0.7472 & 0.7877 \\ 
\hline 
0.3 & 0.9307  & 0.9356 & 0.9336 & 0.9995 \\
\hline
0.35 & 1.1297 & 1.1391 & 1.1364 & 1.2385 \\ 
\hline 
0.4 & 1.3442  & 1.3608 & 1.3581 & 1.5108 \\
\hline
0.45 & 1.5741 & 1.6027 & 1.6019 & 1.8251 \\ 
\hline 
0.5 & 1.8181  & 1.8664 & 1.8718 & 2.1931  \\
\hline
\end{tabular}
\end{center}
\end{table}

   \section{Alternative relations}

The results represented in the previous table suggest the possibility of developing new expressions, allowing us to
describe first dumbbells made of tangent spheres, and then those with overlap.  
So far, relationships between conformal and TPT theories have been presented.  
On the other hand, the contact pair distribution function is known to be better 
described by the expression deduced from the SPT 
than by that obtained with the Ornstein Zernike equation and the PY approximation.
Then, it was first looked for possible alternative relationships between the expression 
of the pressure deduced from 
the WOZ integral equations and the expression of the contact pair distribution function deduced from the SPT.
  
  \subsection{Equations deduced from integral equations}

  In this framework, it was  observed the equality of equation~(\ref{SFTPS}) of the SFP theory 
	with expression~(\ref{PPYb}) obtained independently  with the WOZ integral equations.   
	Then, it seems also interesting to find possible relations between the equations 
	of the TPT  and those deduced from integral equations.   
 The  getting  of the  expression~(\ref{PPYb}) from the equation~(\ref{PPYa}), 
 has been reconsidered, in order to obtain an alternative expression, 
 by assuming that the contact pair distribution function is given by the SPT equation.    
Previously, from equation~(\ref{PPYa}) the relation~(\ref{rhoD}) defining the pair concentration $\rho_{\iD}$ 
 was used to arrive at equation~(\ref{PPYb}).   
  The equality between $g^{00}_{12}(\sigma_{12})$ and~$g^{\iPY}_{12}(\sigma_{12})$ was also used.  
Now, it is assumed that  $g^{00}_{12}(\sigma_{12})$  is equal to $g^{\iSPT}_{12}(\sigma_{12})$  
 in the equation (\ref{rhoD}) defining 
 $\rho_{\iD}$ the concentration of pairs:  
\be
\rho_{\iD } = \rho_1^0 \; \rho_2^0 \; K^0 \; g^{\iSPT}_{12}(\sigma_{12}).
\label{rhoD2}
\ee 
This definition is an improvement because the expression of the pair distribution function at the contact given by SPT is known to be better 
   than  that given by the PY approximation. 
On the other hand,  the equality between   $g^{00}_{12}(\sigma_{12})$ 
 and $g^{\iPY}_{12}(\sigma_{12})$ is  kept   in equation~(\ref{PPYa}), 
 because $g^{00}_{12}(\sigma_{12})$  is  effectively found to be   
 given by the right-hand part of the equation~(\ref{gPY})  in the framework of the WOZ equations.   
  The use of the new definition of the concentration of
 pair $\rho_{\iD}$, then leads, starting from equation~(\ref{PPYa}), by expressing $\rho^0_1 \rho^0_2 K^0$ as a function of~$\rho_{\iD}$, 
 to the alternative expression 
 \be
\beta  P  = \beta  P_{\iPY}^{c} - \rho_{\iD} \; \frac{ \Bigl[ g_{12}^{\iPY}(\sigma_{12})  \Bigr]^2}{ g_{12}^{\iSPT}(\sigma_{12}) }.
 \label{PPYc} 
\ee     
Otherwise, inspired by equation (7.19) 
 seen in~\cite{Rosenfeld88}, the following relation between  $\left[ g^{\iPY}_{12}(\sigma_{12})
 \right]^2$,  $g^{\iSPT}_{12}(\sigma_{12})$  and its derivatives was found
\be 
 g^{\mbox{\rm \tiny SPT}}_{12}(\sigma_{12}) + \sum_k \rho_k \frac{ \partial g^{\mbox{\rm \tiny SPT}}_{12}(\sigma_{12})}
{ \partial \rho_k } = 
 \Bigl[ g^{\mbox{\rm \tiny PY}}_{12}(\sigma_{12}) \Bigr]^2. 
\ee
Therefore, by multiplying by $\rho_1^0 \; \rho_2^0 \; K^0$ on both sides, the following equality is obtained   
\be 
\rho_{\iD} \; \frac{ \Bigl[ g^{\iPY}_{12}(\sigma_{12}) \Bigr]^2 }{ g_{12}^{\iSPT}(\sigma_{12}) }=   
 \rho_{\iD} \; \left[ 1  + \sum_k \rho_k \frac{ \partial 
\ln{ \left[g^{\iSPT}_{12}(\sigma_{12}) \right] }}
{ \partial \rho_k } \right],
\ee
with $\rho_{\iD}$ given by~(\ref{rhoD2}). 
 The left part of the previous equation appears in equation~(\ref{PPYc}) and, 
 in view of equations~(\ref{TPTP}), the right part  is  the relation used in  
  the TPT to calculate the connectivity part of the pressure 
when $g^{\rm ref}(\sigma_{12})$, the reference pair distribution function,  is given by  $g^{\iSPT}_{12}(\sigma_{12})$.    
Therefore, the relation~(\ref{PPYc}), deduced from the WOZ integral equations, 
 can be obtained independently within TPT.  
Since the SPT expression of the contact pair distribution function is better than that given by the PY approximation, 
 the equation~(\ref{PPYc}) is probably an improvement over~(\ref{PPYb}).  
However, the expression of the reference pair distribution at contact with~$g^{\iSPT}_{12}(\sigma_{12})$ 
 is known to be  slightly poorer than that given by the corresponding BM equation. 
Accordingly, the equation~(\ref{PPYc}) is probably somewhat less efficient 
 than the corresponding TPT equation obtained  using the BM approximation.

  \subsection{Equations deduced from conformal theories}

 It has been seen in table~\ref{tbl-ex1}, that, when $l = \sigma_{12}$, the connectivity term given 
 in the expression of the improved free energy ISPT is numerically 
 close to the one found for the initial free energy SPT. The improvement present in the ISPT equations 
 is provided by the 
 first term describing the dissociated hard spheres.  
 The effect of connectivity is taken into account using a simpler equation in the initial SPT theory.  


 In view of these observations, an alternative expression, 
 providing a correction to the SPT expression of the free energy~(\ref{ASPTS}),
  when $l = \sigma_{12}$, was deduced.   
   A modification can be made to the SPT expression by only changing the term of dissociated hard spheres.  
   Then, when $l = \sigma_{12}$, a new equation for  the free energy can be obtained:  
  \be
\beta \frac{A_{\iNew}( l = \sigma_{12})}{V} = \beta \frac{A_{\iBM}}{V} -\rho_{\iD} \ln{{\cal G}_{\iSPT}(\sigma_{12})}.  
\label{AMSPTS}
\ee
Compared with equation~(\ref{ASPTS}), the first term $A^c_{\iPY}$ is replaced by a more 
accurate term $A_{\iBM}$, 
 and the second term remains unchanged because it leads to numerical values close 
 to that given by the corresponding term of the equation~(\ref{AISPTS}).   
The corresponding expression of the pressure is obtained by differentiation of the free energy:  
 \be 
 \beta P_{\iNew} = \beta P_{\iBM} - \rho_{\iD} \; \left[   g_{12}^{\iPY}(\sigma_{12})  -   
  \frac{3 \zeta_2^2}{\left( 1 - \zeta_3 \right)^3 } \;  \frac{\sigma_1^2 \sigma_2^2}{\left( \sigma_1 + \sigma_2 \right)^2 } \right]  
  \quad {\text{for:}} \quad  l = \sigma_{12}.
  \label{MSPTPS}
  \ee 
  This relation allows us to correctly describe the thermodynamics of dumbbells made of tangent spheres. 
As such, it is of little use insofar as many expressions are already available. 
Nevertheless, from this equation a more general relation can be established 
 to describe dimers with overlapping spheres.  
Indeed, equations~(\ref{AMSPTS}) and~(\ref{MSPTPS}) are  linear combinations of various 
relations previously introduced 
\begin{subequations}
\begin{align}
\beta A_{\iNew}  & = \beta A_{\iISFP} + \beta \Bigl[ A_{\iSPT} -  A_{\iSFP} \Bigr],   
\label{AMSPT} \\
\beta P_{\iNew}  & = \beta P_{\iISFP} + \beta \Bigl[  P_{\iSPT} -  P_{\iSFP} \Bigr].  
\label{PMSPT} 
\end{align} 
\end{subequations} 
 \begin{figure}[!t]
\centerline{\includegraphics[width=0.70\textwidth]{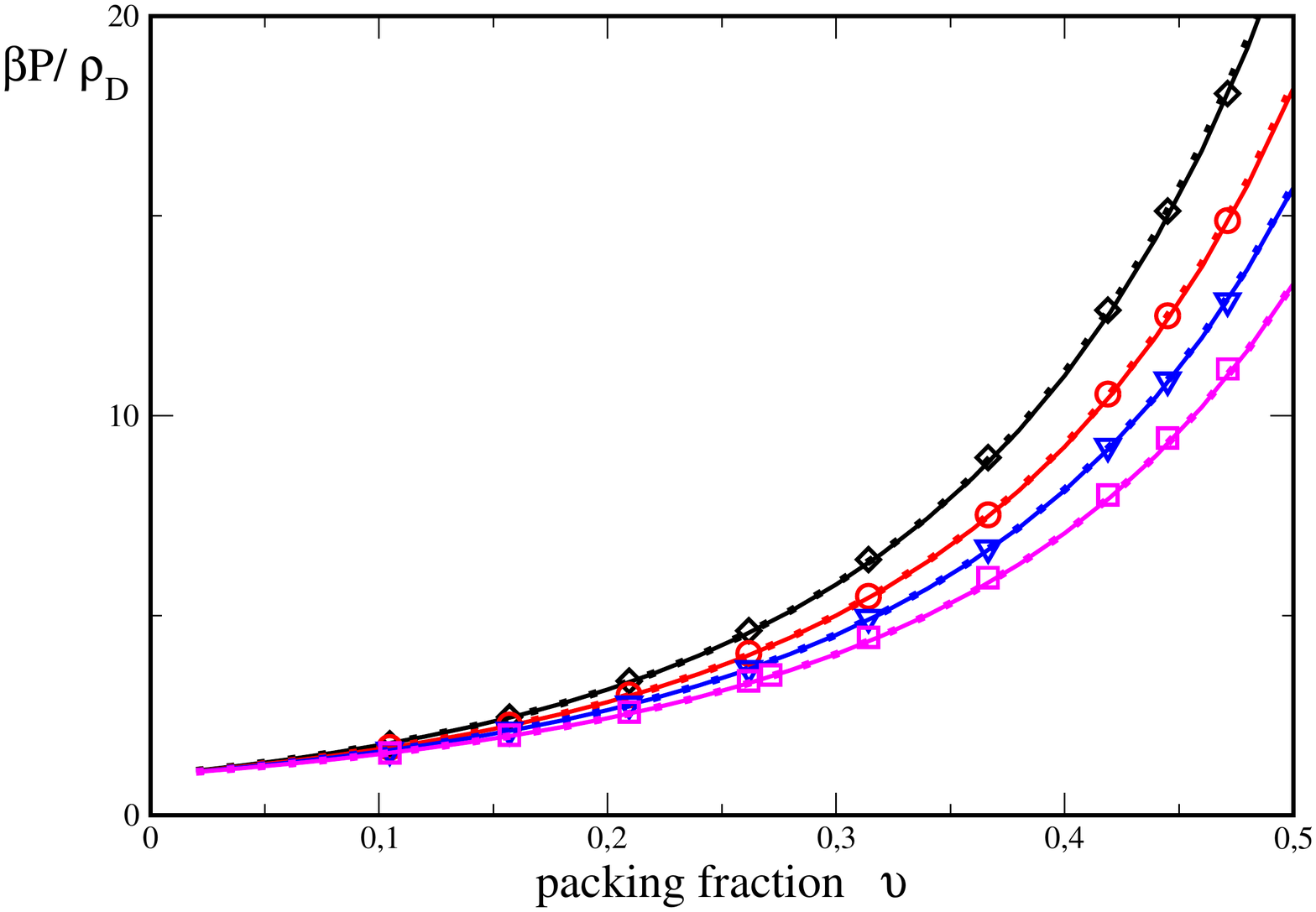}}
\caption{(Colour online) Equation of state for overlapping hard sphere dumbbells. 
 From top to bottom, lines represent  theoretical results for different overlap distances:        
$\cdot  \cdot  \cdot \, \cdot$  New equation (\ref{PM2}),  --- ISPT equation (\ref{ISPTP}). Symbols are the simulation results 
\cite{Tildesley80}.    
Diamonds and black lines: $l=1$; Circles and red lines: $l=0.8$; Triangles and blue lines: $l=0.6$, 
Squares and pink lines: $l=0.2$. } \label{fig-ex2}
\end{figure} 
These new relationships add corrective terms to the previous ISFP equations.  
Moreover, these  latter expressions can also be used for dimers with overlapping of the spheres 
by using the more general equations introduced previously to describe them.       
Therefore, using equations~(\ref{ISFPP}), (\ref{SPTP}) and~(\ref{SFTP}), the   
 equation~(\ref{PMSPT}) has been evaluated for a dumbbell fluid 
 where the distance $l$ between sphere is less than or equal to $1$.  
 The term in square brackets, $\left[ P_{\iSPT} - P_{\iSFP} \right]$ in the previous equation, can be simplified, giving 
 \be 
 \beta P_{\iNew}    = \beta P_{\iISFP} +  \frac{s^2}{3 \left( 1 - v \right)^3} \;  \left( q - \frac{s}{4 \piup} \right),
 \label{PM2}
 \ee  
  where the difference $q - s / (4 \pi)$, starting from equations~(\ref{Sdb}) and~(\ref{Rdb}), can be evaluated according to
 \be
  \left( q - \frac{s}{4 \piup} \right) = \frac{\rho_{\iD} }{16} \left[ l - \frac{\left(\sigma_1 - \sigma_2 \right)^2}{ 4 l } \right]^2.
  \ee
 The results of the calculations performed with equation~(\ref{PM2}) are presented in figure~\ref{fig-ex2}, 
for  homonuclear dimers, for various  separations of the constituents.  
A comparison is also given with the results deduced from the ISPT equation~(\ref{ISPTP}).  
It is observed that the new theoretical expression describes well the values obtained by simulation.  
 It provides  values numerically very close to those given by the ISPT equation.  
The new equation can be seen also as an improvement of the improved SFP equation~(\ref{ISFPP}).    
The corrective term for $P_{\iISFP}$, presented in equation~(\ref{PM2}), seems to correspond numerically 
to the contribution required to recover the results obtained with the expression of $P_{\iISPT}$.

\section{Conclusion}

In the previous sections, it has been shown how, from the conformal theories
 applied to dumbbells made of tangent hard spheres, the pressure and the free energy can be split 
 into a term of repulsion
of hard spheres and a term of connectivity between these spheres.  
This separation is similar to that encountered in the expression of pressure in the TPT theory.  
The analysis presented in this work  highlights the relative influence of these two contributions 
in the different expressions presented in various conformal theories studied.   
 It has been found that  for each class of conformal
theory the reference term is, for the original version that given by the PY theory, and
for the improved version that given by the BMCSL expression. Then, when the scaled
particle and the scaled field theory are compared, it was noted that the terms of reference 
 are the same, both for the original approach and for the improved expressions. On
the other hand, for each of the two classes of conformal theory considered, the term of
connectivity changes little when one goes from the original expression to the improved
one. In addition, the term of connectivity is different between the scaled field and the
scaled particle theory.  
Connections and similarities have also been seen between the expressions describing these two contributions 
and those found  with the TPT for dimers  made of tangent spheres.  
Similarities were also found with some relationships given in the context of integral equations. 
  This work allows one to explain differences found in  the description of 
 thermodynamic properties deduced from simulations.

Furthermore, this analysis has allowed  one to provide alternative expressions of the connectivity terms, 
in particular for dimers with overlap. 
 It could be extended to fluids containing chains of a greater number of constituents.  
It could also be useful to better represent subunits belonging to polymers according to the group 
 contribution formalism~\cite{Lymperiadis08}.  
 On the other hand, the analysis of thermodynamic quantities such as free energy and pressure could 
 be extended to the activity coefficients of  various species present in the fluids of dimers~\cite{Labik95}.    
Then, the study of the chemical equilibrium between the hard spheres forming the overlapping dimers 
 could be reviewed. Thus, this will allow one to describe the background correlation in continuation of 
 previous studies on this subject~\cite{Boublik86,Labik88,Boublik06}.

\ukrainianpart

\title{Взаємозв'язок між термодинамічними збуреннями та теоріями масштабованої частинки для конденсованих рідких димерів
}
\author[О. Бернар]{О. Бернар}
\address{Лабораторія PHENIX, CNRS, Університет Сорбонни (Кампус П'єра та Марії Кюрі), 4 Плас Жюссіє, F-75005 Париж, Франція}

\makeukrtitle

\begin{abstract}
	У роботі зроблено огляд різних підходів для побудови теорії масштабованої частинки з метою опису  плинів гантелеподібних частинок, утворених з твердих сфер, що перекриваються або торкаються одна одної. 
	Надано інше представлення відомим з літератури виразам у математичній формі, подібній до тієї, яка відома з термодинамічної теорії збурень, отриманої Вертгаймом для ланцюжків і узагальненої у статистичній теорії асоціативних рідин. Аналогії та відмінності між цими двома теоретичними підходами дозволили запропонувати альтернативні вирази для опису плинів гантелеподібних частинок зі сферами, що перекриваються.
	\keywords  плини гантелеподібних частинок з твердими серцевинами, теорія масштабованої частинки, термодинамічна теорія збурень

\end{abstract}
\lastpage
\end{document}